\documentclass{article}

\usepackage{PRIMEarxiv}
\usepackage{subcaption}
\usepackage[utf8]{inputenc} % allow utf-8 input
\usepackage[T1]{fontenc}    % use 8-bit T1 fonts
\usepackage{hyperref}       % hyperlinks
\usepackage{url}            % simple URL typesetting
\usepackage{booktabs}       % professional-quality tables
\usepackage{amsfonts}       % blackboard math symbols
\usepackage{nicefrac}       % compact symbols for 1/2, etc.
\usepackage{microtype}      % microtypography
\usepackage{lipsum}
\usepackage{fancyhdr}       % header
\usepackage{graphicx}       % graphics
\usepackage{longtable}

\graphicspath{{media/}}     % organize your images and other figures under media/ folder

\title{User Perceptions of Automatic Fake News Detection: Can Algorithms Fight Online Misinformation?}

\author{
  Bruno Tafur \\
  University of Cambridge \\
  Cambridge \\
  \texttt{bt403@cantab.ac.uk} \\
   \And
  Advait Sarkar \\
  University of Cambridge \\
  Cambridge \\
  \texttt{advait@microsoft.com} \\
}

\begin{document}
\maketitle

\begin{abstract}
    Fake news detection algorithms apply machine learning to various news attributes and their relationships. However, their success is usually evaluated based on how the algorithm performs on a static benchmark, independent of real users. On the other hand, studies of user trust in fake news has identified relevant factors such as the user's previous beliefs, the article format, and the source's reputation. We present a user study (n=40) evaluating how warnings issued by fake news detection algorithms affect the user's ability to detect misinformation. We find that such warnings strongly influence users' perception of the truth, that even a moderately accurate classifier can improve overall user accuracy, and that users tend to be biased towards agreeing with the algorithm, even when it is incorrect.
\end{abstract}

% keywords can be removed
\keywords{fake news, machine learning, misinformation, user trust}

\section{Introduction}
False information can be rapidly published and spread online, reaching massive audiences \cite{zhou_survey_2020, thorne_automated_2018, lasser2022social}. Social networks have catalyzed the spread of misinformation, affecting people worldwide. In response, research has focused on fake news detection using machine learning \cite{wang_liar_2017, antoun_state_2020, sepulveda-torres_exploring_2021,mazzeo_detection_2021,aldwairi_detecting_2018}. These algorithms typically rely on analysis of content, information source, or propagation patterns \cite{antoun_state_2020}. High accuracy models use collections of features \cite{antoun_state_2020, zhang_fakedetector_2020, shu_beyond_2019}.

Studies have measured users' ability to detect fake news \cite{vziatysheva_testing_2021}, and users' change in trust with respect to changes in user and expert reputation ratings or changes in content structure and format  \cite{kim_says_2018,kim_combating_2019, luo_credibility_2020, spezzano_thats_2021,kirchner_countering_2020}. Researchers have proposed behavioural ``nudges'', causing users to distrust or critically evaluate news before sharing \cite{lamas_combating_2019, jahanbakhsh_exploring_2021}. Such nudges could be generated by a fake news detection algorithm, or reputation ratings.

Some studies have evaluated the use of warnings in news articles, generally focusing on a subset of fake or disputed articles. \cite{kirchner_countering_2020, pennycook_implied_2020,clayton_real_2020}. The focus is mostly on analysing the effect of textual information and the design of the articles when showing the warnings.

However, the user perspective and interaction with fake news detection algorithms is under-explored. What might it be like for a user to browse the web, a mixture of true and false information, with accuracy ratings issued by an algorithm? Moreover, what if the algorithm, as is usually the case, makes mistakes?

%Analysing these aspects can bring new insights into how the user reacts to being warned by a system and its impact when the system is correct or wrong. Therefore, it would be useful to evaluate how trustworthy these detection algorithms are from an end user's perspective and which factors are involved in the process.

% This experiment was conducted via a survey based on a fake news dataset of 40 news articles and answered by 40 participants.

In this paper, we explore user interaction with fake news detection algorithms, by measuring how frequently users agree with algorithm warnings in different scenarios. We evaluate if an imperfect algorithm, on the whole, can help users better identify misinformation. We make the following contributions:
\begin{itemize}
  \item We present an empirical study with 40 participants, demonstrating that using fake news detection algorithms can improve user ability to discern fake news from real news.
  \item We demonstrate that user \emph{accuracy} when detecting fake and real news can vary depending on the truthfulness of the article and the algorithm's correctness.
  \item We present evidence that users' \emph{agreement} with a fake news algorithm is affected by its correctness, and that because users tend to agree with system advice, incorrect warnings may be worse than no warnings.
  \item We confirm previous results that suggest that users' accuracy in identifying fake and real news is relatively low \cite{kirchner_countering_2020, luo_credibility_2020, snijders2022humans}. Specifically, in our case, participants had 61\% accuracy without algorithm support. 
\end{itemize}

\section{Related Work}

Discussions of fake news are made complex due to its various definitions \cite{zhou_survey_2020}. Related topics include misinformation and disinformation. Misinformation refers to distributed information that is false or inaccurate, while disinformation is the distribution of misinformation on purpose with a deceptive intent \cite{thorne_automated_2018}. Fake news can include concepts such as rumours, satire and disinformation \cite{zhou_survey_2020}. In our work, we make the simplifying assumption that fake news is articles with false factual statements, where falsity is determined by a label in our dataset (which in turn derives from the perceptions of data annotators, who attempt to reflect scientific consensus), regardless of the intent of the publisher.

\subsection{Fake news detection algorithms}
Fake news detection algorithms apply supervised machine learning methods, commonly based on content analysis, source analysis or propagation patterns \cite{antoun_state_2020}. Propagation-based models often use graph neural networks \cite{bian_rumor_2020, han_graph_2020, monti_fake_2019}, analysing the patterns of news spread. Other models focus on content, classifying based on the style of the text or comparison with a factual knowledge base \cite{han_graph_2020, perez-rosas_automatic_2018,wang_liar_2017,afroz_detecting_2012}.

%  challenge is generalising such algorithms to unseen networks \cite{han_graph_2020}.

The most effective models have used mixed approaches \cite{zhang_fakedetector_2020, antoun_state_2020, shu_beyond_2019}. For example, Antoun et al. \cite{antoun_state_2020} use features from the headline, the body and the relationship with the top five searches in Google. Their tests employ transformer architectures, such as RoBERTa and XLNet, and reached an F1-score of up to 98\% when detecting fake news with data from the Qatar International Cybersecurity Contest. The model developed by Shu et al. \cite{shu_beyond_2019} achieved an 88\% F1-score in Politifact data and 87\% in Buzzfeed data. These studies focus on the performance of algorithms on static benchmarks, and not on the user experience.

% Their method examines the content and also examines the relationship to other articles.

\subsection{User studies}
Research has measured users' detection accuracy of fake news, taking into consideration factors such as news format, topic, structure and reputation \cite{kim_says_2018,kim_combating_2019, luo_credibility_2020, spezzano_thats_2021,kirchner_countering_2020}. There is a diversity of terminology and measurements: studies refer variously to user trust, believability, credibility or user engagement \cite{vziatysheva_testing_2021}.

Kim \& Dennis \cite{kim_says_2018} analysed how changing the presentation format of an article affects users' trust and further actions on social media, finding that showing the source of an article increased scepticism and having low source ratings affected the article's believability. Similarly, Kim et al. \cite{kim_combating_2019} analysed the impact of user, source and expert reputation ratings on article credibility. They found that all three kinds of ratings influenced trust. Both studies found confirmation bias: users tend to believe more in articles aligned with their previous beliefs.

A study by Spezzano et al. \cite{spezzano_thats_2021} examined how having an image, title, excerpt, and source affected user's trust. The study then compared human-level detection to the performance of an algorithm. Human accuracy results showed a 62\% detection accuracy when title and image were included versus 53\% when they were just exposed to an extract, but their detection algorithm achieved 83\% accuracy.

Another study analysed how credibility was affected by the topic of the news and by the number of likes on social media \cite{luo_credibility_2020}. Their results showed an average user detection accuracy of 51\%, and users had a higher detection accuracy with fake news than real news. The authors suggested that this difference could be due to a deception-bias, i.e., a presumption of deception by the users. They also found a relationship between Facebook likes and the higher credibility.

Some studies have evaluated applying warnings to articles \cite{kirchner_countering_2020, pennycook_implied_2020, clayton_real_2020, seo_trust_2019, snijders2022humans} or informational checklists \cite{heuer_comparative_2022}. Kirchner \& Reuter \cite{kirchner_countering_2020} tested warnings on fake news, finding that users prefer to be warned about fake news and that the warnings affected the perceived accuracy of false headlines. Similarly,  Pennycook et al. \cite{pennycook_implied_2020} showed that applying warnings to a subset of fake news impacts the perceived accuracy of fake news headlines. However, the articles that did not show a warning were negatively affected as users tend to assume that the absence of a warning made the news article more truthful. This is interpreted as an ``implied truth'' effect and suggests warnings should be balanced between fake and real news. Also, Clayton et al. \cite{clayton_real_2020} found that using "Rated False" and "Disputed" tags reduced the belief in fake news, with "Rated False" having a more significant effect. Therefore, there could also be an impact generated by the type of warning used.

Seo et al. \cite{seo_trust_2019} evaluated effects of 3 types of warnings on participants' recognition, detection and sharing of fake news. They found that giving explanations inside the warning had a positive effect on the user's decision. The study did not cover possible scenarios of algorithm incorrectness and its impact on users' perception of the truth, and focused on adding warnings to disputed articles but not to non-disputed ones.

A close precedent to our work is Snijders et al. \cite{snijders2022humans}, who explore effects of individual confidence on trust in the advice of an algorithmic fake news detector, finding that participants are less likely to accept algorithmic advice for news stories about which they are themselves confident. Crucially, the algorithm used in their study was not fully accurate. Another study by Lu et al. \cite{lu_effects_2022} also analysed nudges by an AI algorithm and analysed the effect related to news spread. However, in both studies, their report does not differentially analyse user agreement in the cases where the algorithmic advice was correct versus erroneous.

\textbf{In summary}, previous research has developed accurate, yet imperfect algorithms for fake news detection. Studies of users have shown that cues such as format, topic, prior beliefs, source, author, images, etc. all contribute to the perceived credibility of a news article. Furthermore, studies have shown that warnings can help users evaluate misinformation. However, no previous user study has contrasted the full set of scenarios that may arise with an imperfect fake news detector in practice, including true information that has been incorrectly flagged as false, or false information that has been incorrectly flagged as true. Our study fills this gap.

\section{Method}

We designed a study to investigate the impact of an imperfect fake news detection algorithm on users' ability to detect fake and real news.

\paragraph{Participants.} We recruited 40 English-speaking adults via the Amazon Mechanical Turk platform. We selected participants from English-speaking countries, who had carried out more than 1,000 surveys on the platform, and with an approval rate of more than 99\%.

\paragraph{Dataset.} We manually curated a dataset 40 news articles.\footnote{Archived on the Open Science Framework at \url{https://osf.io/bjn2q/}} Twenty articles were compiled manually from recent publications in Snopes and Politifact, fact-checking websites that have been used in previous research to generate similar datasets \cite{torabi_asr_big_2019, torabi_asr_data_2018, shu_fakenewsnet_2020}. A further 10 articles were picked from the MisInfoText dataset \cite{torabi_asr_data_2018,torabi_asr_big_2019} and a final 10 articles were selected from the Fake and Real dataset \cite{ahmed_detecting_2018, ahmed_detection_2017}. Each article consisted of a headline and a single paragraph of content. We chose this short length to reduce reading time, so that participants could be exposed to a large number of articles during the study. We selected articles based on the following criteria:
\begin{itemize}
    \item Articles needed to be easily established as true or false based on publicly available information from authorative sources. We removed articles where we could not determine truth or falsity with high certainty. We acknowledge that this is a subjective step.
    \item Articles needed to be current at the time of the study.
    \item The dataset needed to be balanced, having the same number of true and false articles, and having a diverse range of topics that were roughly equally represented. We chose articles representing the following broad topics: food \& health, politics, climate change, world news, and COVID-19.
\end{itemize}

% Our dataset is published in Open Science Framework and can be found in https://osf.io/bjn2q/.
%The main focus was having the same amount of fake and real news, having an even balance of topics, and choosing topics that can be considered relevant during the time of the study. Therefore, the following 

\begin{figure}[t]
    \centering
    \begin{subfigure}[b]{0.45\textwidth}
        \includegraphics[width=\textwidth]{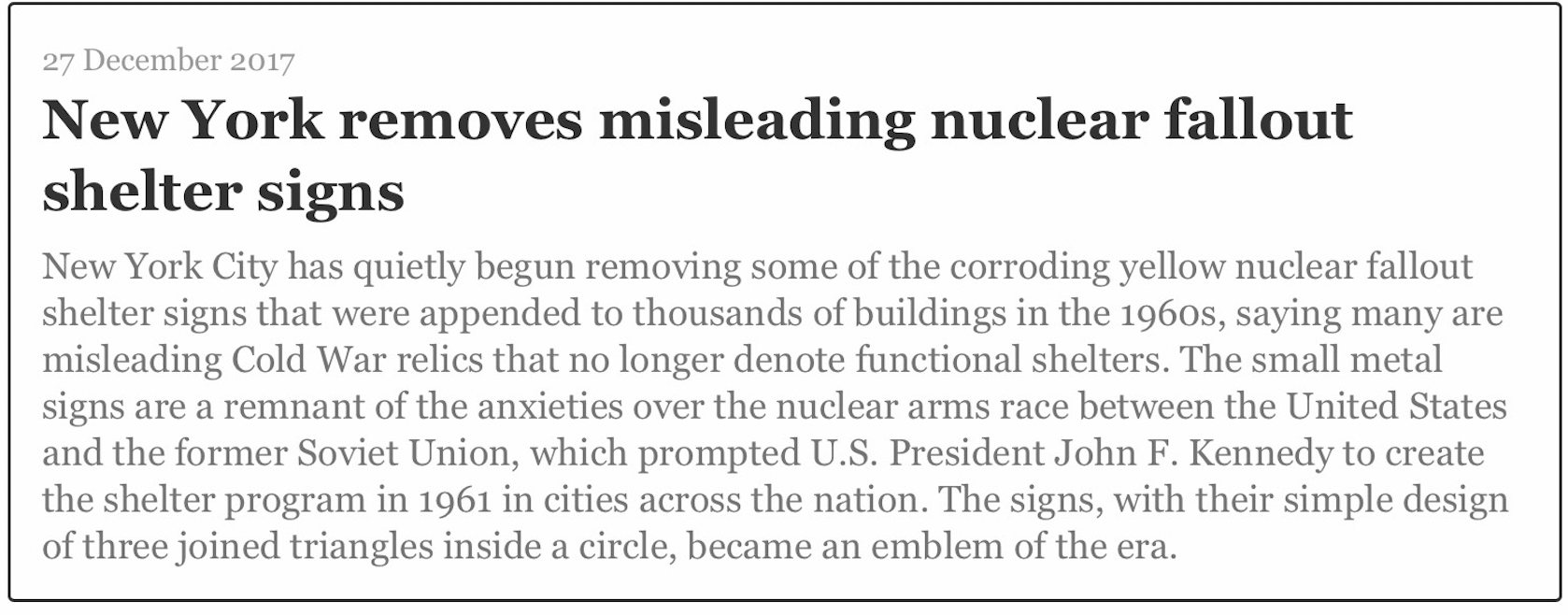}
        \caption{Without warning}
        \label{fig:without-warning}
    \end{subfigure}
    \begin{subfigure}[b]{0.45\textwidth}
        \includegraphics[width=\textwidth]{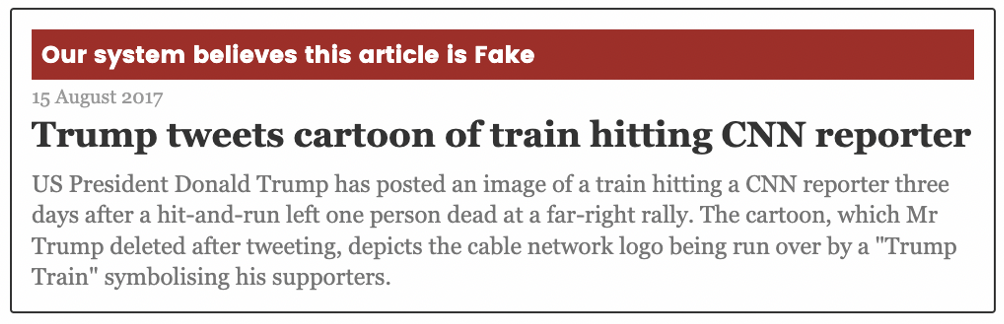}
        \caption{With fake article warning}
        \label{fig:with-fake-warning}
    \end{subfigure}
    \break

    \begin{subfigure}[b]{0.45\textwidth}
        \includegraphics[width=\textwidth]{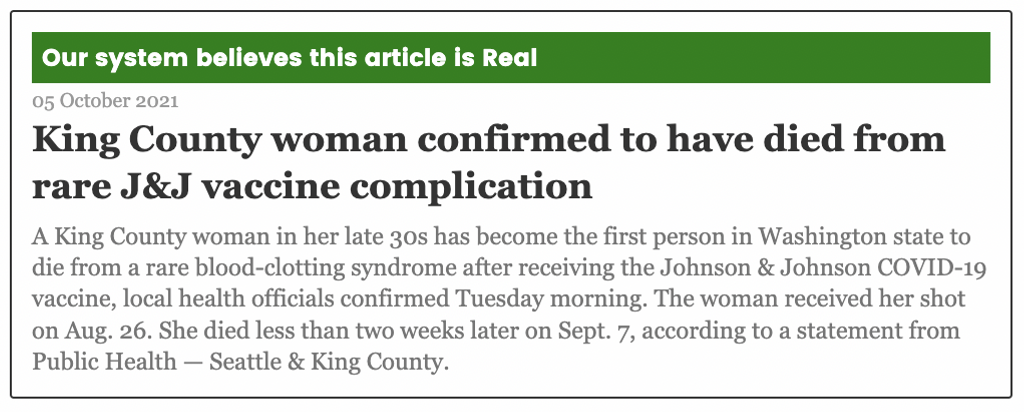}
        \caption{With real article warning}
        \label{fig:with-real-warning}
    \end{subfigure}
    \begin{subfigure}[b]{0.45\textwidth}
        \includegraphics[width=\textwidth]{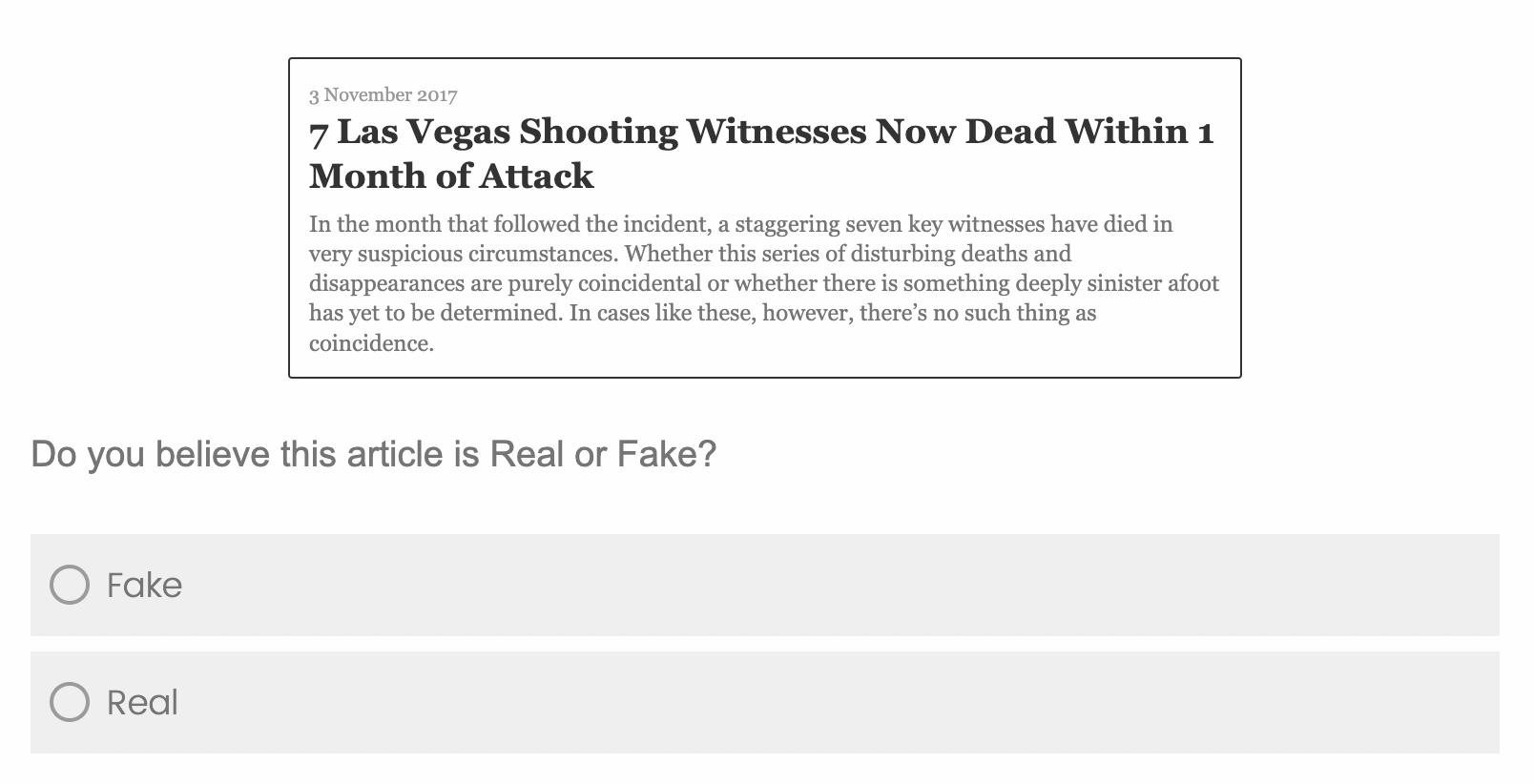}
        \caption{Sample question shown in the survey (no warning shown)}
        \label{fig:with-real-warning}
    \end{subfigure}
    \caption{Sample news articles shown in the survey}\label{fig:sample-news-articles}
\end{figure}

\paragraph{Protocol.} We adopted a counterbalanced, within-subjects design. All participants were shown 40 articles, and asked to rate them as fake or real. Each user rated half of the articles with algorithm support (with warnings) and half of the articles without warnings. The articles with algorithm support displayed a warning based on the classification output of a (hypothetical) fake news detection algorithm with imperfect accuracy. Examples of the articles with and without warnings, and an example of a full question shown in the survey, can be seen in Figure~\ref{fig:sample-news-articles}. Sample headlines for each topic can be seen in Table~\ref{tab:sample-headline}.

% For purposes of this study, our fake news detection algorithm will be simulated by the use of warnings, following what a real algorithm could show

To counterbalance the stimuli used in the with and without-warnings conditions, the 40 articles were divided into two batches of twenty articles, containing 10 real and 10 fake articles each. The 40 participants were randomly divided into two equal groups. Group A of the participants was shown Batch 1 without warnings and Batch 2 with the algorithm detection warnings. Group B was exposed to Batch 1 with warnings and Batch 2 without warnings. This is illustrated in Table~\ref{tab:groups}. For each participant, the items within Batch 1 and Batch 2 were randomly shuffled, to mitigate order effects.

\begin{table}[t]
  \caption{Participant groups and condition assignments}
  \label{tab:groups}
  \centering
  \begin{tabular}{l l}
    \toprule
    \bf{Participant group A}  & \bf{Participant group B}\\
    \midrule
    Batch 1, without warnings   & Batch 1, with warnings \\
    Batch 2, with warnings      & Batch 2, without warnings \\
  \bottomrule
\end{tabular}
\end{table}

\begin{table}[t]
\caption{Sample article headlines per topic.}
\label{tab:sample-headline}
\centering
\begin{tabular}{l|p{0.35\textwidth}|p{0.35\textwidth}}
\toprule
Category       & Fake                                                                                                 & True                                                              \\ \midrule
Food \& Health & FBI Issues Horrifying Warning to Frequent Grocery Shoppers                                           & ``Rattlesnake selfie'' results in a \$153K medical bill             \\
Politics       & Trump Just Got Banned From The Place He Proposed To Melania                                          & Trump tweets cartoon of train hitting CNN reporter                \\
Climate Change & Facebook Spamming Climate Posts with ``Climate Science Center'' Propaganda                             & Amid higher global temperatures, sea ice at record lows at poles  \\
World          & Billion Dollar Company Tells Employees How They’re Allowed To React On Social Media To 46\% Pay Cuts & Nigeria says U.S. agrees delayed \$593 million fighter plane sale \\
COVID          & 1000 Peer Reviewed Studies Questioning Covid-19 Vaccine Safety                                       & Trump booed after revealing he got a Covid booster shot           \\ 
\bottomrule
\end{tabular}

\end{table}

% \begin{figure}[t]
%   \centering
%   \includegraphics[width=0.9\textwidth]{sample-articles-per-topic.png}
%   \caption{Sample article headlines per topic.}
%   \label{fig:sample-headline}
% \end{figure}

%The articles showed a heading, the date and an extract to the user, as seen in Figure 1. 

In practice, fake news detection algorithms report accuracies between 60\% and 90\% \cite{shu_beyond_2019, han_graph_2020}. We therefore chose to make our simulated algorithm 70\% accurate, aligning with previous work \cite{snijders2022humans}. Thus, in 30\% of the cases, the algorithm either incorrectly flagged a true article as false, or a false article as true. In practice we randomly selected 3 (out of 10) real and 3 fake stories per batch to show incorrect warnings for; each participant thus encountered 14 correct and 6 incorrect warnings in the with-warnings condition. This falls in the range of possible accuracies of a real algorithm, and allows sufficient instances of misclassification to observe interesting data about user responses in that range.

The study did not collect any personal or identifiable information, and was approved by our institutional ethics committee. 

% Therefore, this setup allows experimenting with the impact of having mistakes in the warnings supplied to the user as well as having a more realistic setup as no algorithm is 100\% accurate. 

\section{Results}
% The 40 participants rated 40 articles as fake or real. Half of these articles showed a warning based on what the fake news detection algorithm would show. The algorithm had an accuracy of 70\%, so 30\% of the warnings were false.

We define user \emph{accuracy} as the percentage of articles correctly classified by the user. We define user \emph{agreement} with the algorithm as the percentage of articles where the user rating (real or fake) matches the algorithm rating.

% measured by how frequently the users follow the algorithm instructions.

\subsection{User accuracy}

\subsubsection{Effects of algorithm warnings} %\hspace*{\fill}

% \begin{table}[t]
%   \caption{User accuracy with and without fake news detection algorithm warnings}
%   \label{tab:accuracy-with-without-warnings}
%   \begin{tabular}{c c c c} 
%     \toprule
%     \bf{Batch} & \bf{Acc. With Alg. Warning ($\sigma$)} & \bf{Acc. With No Alg. Warning ($\sigma$)} & \bf{Mean Acc.} \\
%     \midrule
%     Batch 1 & 69.0\% ($\pm$ 8.0\%) & 57.8\% ($\pm$ 14.6\%) & 63.4\% \\
%     Batch 2 & 67.0\% ($\pm$ 10.8\%) & 63.3\% ($\pm$ 11.6\%) & 65.1\% \\
%     \midrule
%     \bf{Mean Acc.} & \bf{68.0\% ($\pm$ 9.5\%)} & \bf{60.5\% ($\pm$ 13.2\%)} & \\
%   \bottomrule
% \end{tabular}
% \end{table}

%Both sets of accuracy results were tested for normality using the Shapiro-Wilk test, obtaining a p-value of 0.096 with algorithm warnings and a p-value of 0.056 without algorithm warnings, both greater than 0.05 and rejecting the hypothesis that they are not normally distributed. 

% Figure 4 shows the distributions as a graph. Additionally, the t-test was carried out to determine that these results could not have happened at random. From the t-test, the p-value obtained was ${7.78 x 10^{-12}}$, which is considerably lower than 1\%. Therefore, the distributions are considered significantly different. Finally, the mean difference was 7.5 percentage points, and Cohen's d was calculated to quantify the standardized effect size of the difference, obtaining a value of 0.649, which can be considered a medium to large effect.

\begin{figure}[t]
  \centering
  \includegraphics[width=0.7\textwidth]{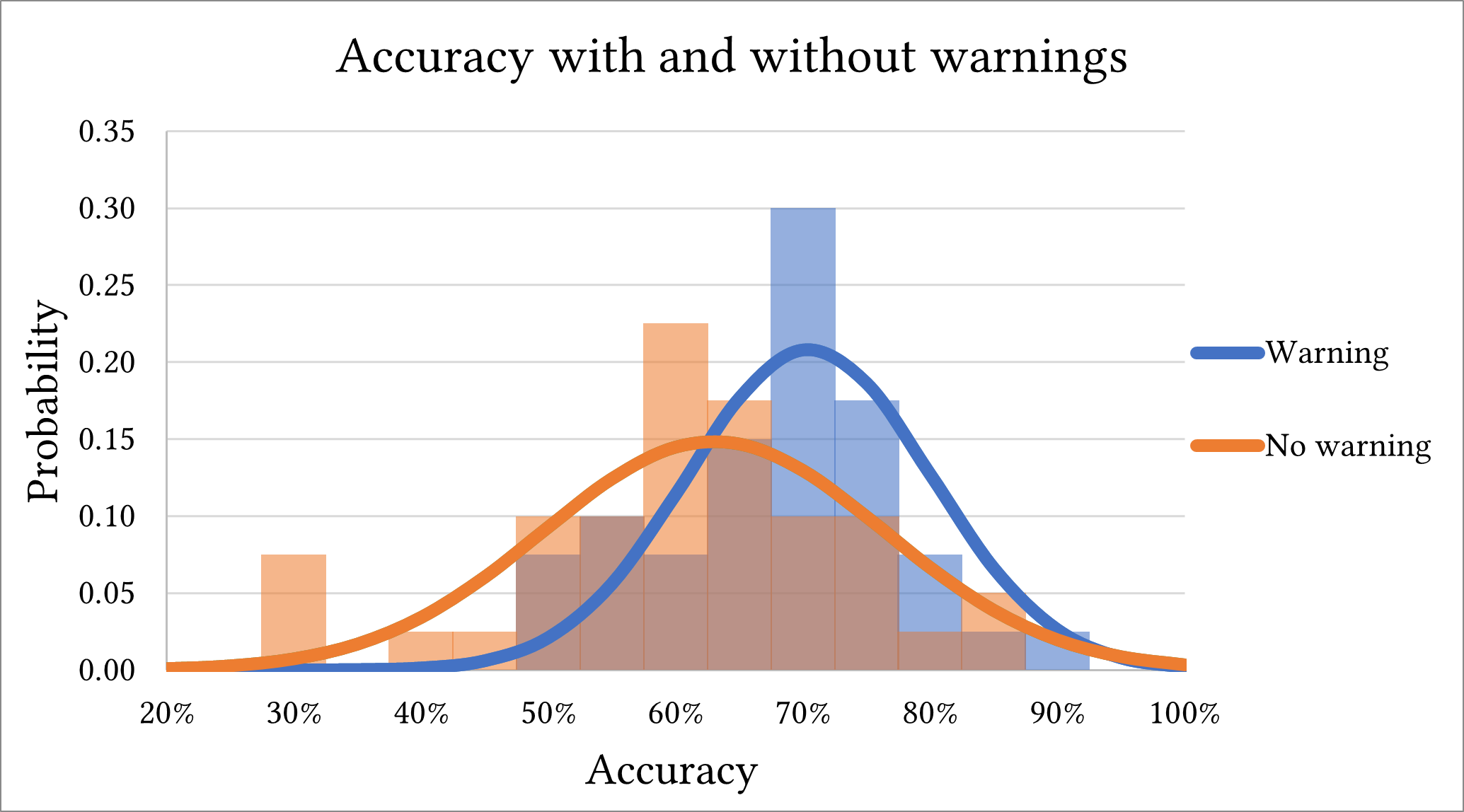}
  \caption{Normalised distribution of user accuracy with and without algorithm warnings. Accuracy improves with warnings.}
  \label{fig:accuracy-distributions}
\end{figure}

The mean user accuracy with and without warnings is shown in Table~\ref{tab:accuracy-fake-real}. Accuracy increased from 60.5\% without warnings to 68\% with warnings. This difference is a medium effect size (Cohen's D $=0.65$) and is statistically significant (t-test, $p=7.78 \cdot 10^{-12}$). The distributions can be seen in Figure~\ref{fig:accuracy-distributions}.

\subsubsection{Effects of truth, falsity, and algorithm correctness}

% Similarly, the results were also broken down to analyse if the nature of being fake or real generated a significant accuracy difference. To illustrate this,

\begin{table}[t]
  \centering
  \caption{User accuracy for fake and real news, with and without algorithm warnings}
  \label{tab:accuracy-fake-real}
  \begin{tabular}{c c c c} 
    \toprule
     & \bf{Acc. Fake ($\sigma$)} & \bf{Acc. Real ($\sigma$)} & \bf{Mean Acc. ($\sigma$)}  \\
    \midrule
    \bf{With warning} & \bf{75.3\% ($\pm$ 13.5\%)} & \bf{60.8\% ($\pm$ 15.2\%)} & \bf{68.0\% ($\pm$ 9.5\%)}\\
    Correct warning & 85.7\% & 72.1\% & 78.9\%\\
    Incorrect warning & 50.8\% & 34.2\% & 42.5\%\\
    \midrule
    \bf{Without warning} & \bf{63.0\% ($\pm$ 20.5\%)} & \bf{58.0\% ($\pm$ 17.2\%)} & \bf{60.5\% ($\pm$ 13.2\%)}\\
    \midrule
   \bf{Overall mean} & \bf{69.1\% ($\pm$ 15.1\%)} & \bf{59.4\% ($\pm$ 12.5\%)}  \\
  \bottomrule
\end{tabular}
\end{table}

Table~\ref{tab:accuracy-fake-real} breaks down user accuracy by fake and real articles. The average user accuracy for detecting fake news is 69\%, while for real news it is 59\%. This difference is statistically significant (Wilcoxon signed-rank test, $z=2.75, p=0.006$. A non-parametric test is used here as the data are not normally distributed).

Warnings in Table~\ref{tab:accuracy-fake-real} are subdivided into correct and incorrect. User accuracy with correct warnings was considerably higher than with incorrect warnings, going from 42.5\% to 78.9\%. This difference is statistically significant (Wilcoxon signed-rank test, $z=4.14, p=4 \cdot 10^{-5}$).

\subsubsection{Effect of news topics}

\begin{figure}[t]
  \centering
  \includegraphics[width=\textwidth]{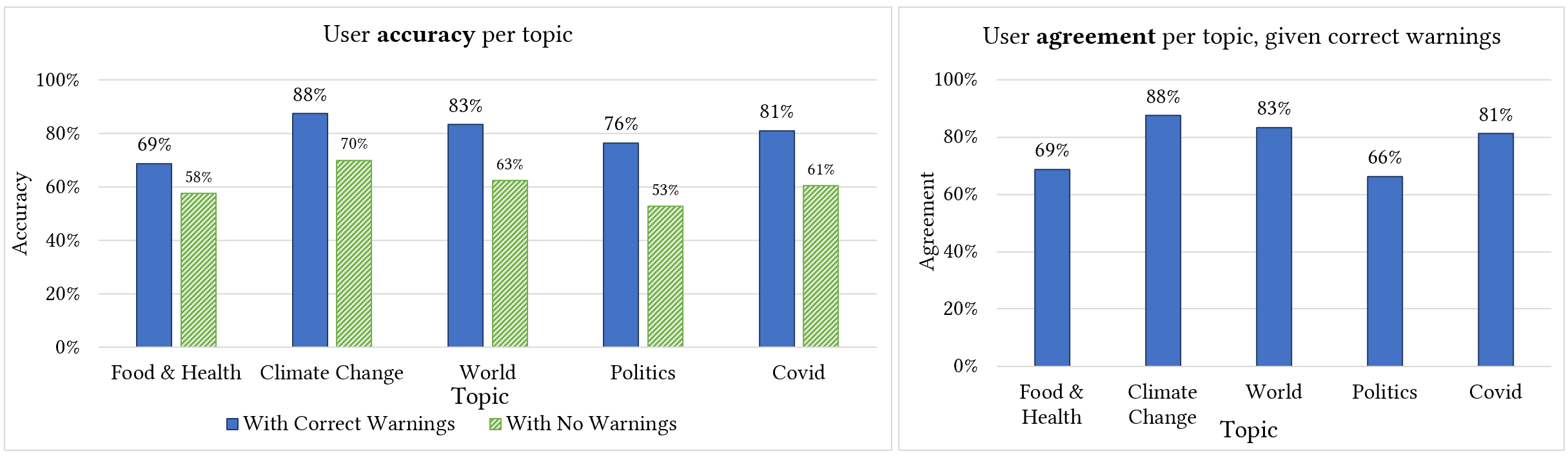}
  \caption{Left: Comparing user accuracy per topic with correct warnings, versus without warnings. Right: user agreement with algorithm per topic, with correct warnings.}
  \label{fig:topic-accuracy-agreement}
\end{figure}

%Additionally, this study analysed the accuracy in specific news topics to see how this might have influenced the experiments. These results can be seen in Figure 5. 
Figure~\ref{fig:topic-accuracy-agreement} (left) shows user accuracy by topic, restricted just trials where there was no warning, or a correct warning (our sample was not large enough to investigate incorrect warnings by topics, which we leave for future work). The difference between topics is statistically significant (Friedman test, $p=0.007$), but this topic needs further investigation as the number of news articles in each topic is relatively small. We include this preliminary analysis here as a starting point for such investigation.

\subsection{User agreement with the algorithm}

We examined the frequency with which users agreed with the algorithm ratings, whether fake or real. On average, the users agreed with the algorithm decision 72.5\% of the time.
%Table 4 shows a segmented view of the results of the users compared with the system and Table 5 shows the segmented percentages of agreement.

% We can loosely interpret this agreement as an indicator of the user's trust in the algorithm.

% \subsubsection{Effects of fake vs. real news}\hspace*{\fill}

Table~\ref{tab:agreement} summarizes user agreement with the warnings, when the underlying article is fake or real, and the warning is correct or wrong. Users agreed with the algorithm 74.8\% of the time for fake articles and 70.3\% of the time for real articles, but this difference is not statistically significant. Nonetheless, this can be loosely interpreted as the users showing slightly more trust in the algorithm when the warnings were related to fake news.

\begin{table}
\centering
  \caption{Fake vs. real news user agreement with and without algorithm warnings}
  \label{tab:agreement}
  \begin{tabular}{c c c c} 
    \toprule
    & \bf{Agreement Fake ($\sigma$)} & \bf{Agreement Real ($\sigma$)} & \bf{Mean Agreement ($\sigma$)} \\
    \midrule
    Correct warnings & 85.7\% & 72.1\% & 78.9\% ($\pm$ 15.7\%) \\
    Incorrect warnings & 49.2\% & 65.8\% & 57.8\% ($\pm$ 31.3\%)\\
    \midrule
    \bf{Mean Agreement} & \bf{74.8\% ($\pm$ 18.0\%)} & \bf{70.3\% ($\pm$ 23.1\%)} & \bf{72.5\% ($\pm$ 18.1\%)}\\
  \bottomrule
\end{tabular}
\end{table}

% \subsubsection{Effects of algorithm correctness}\hspace*{\fill}

The users agreed with the algorithm recommendations 78.9\% of the time when the algorithm was correct, and the users agreed with the algorithm 57.8\% of the time when it was incorrect. This difference is statistically significant (Wilcoxon signed-rank test, $z=4.17, p=10\cdot 10^{-5}$). Thus, users are considerably more likely to agree with the algorithm when it is giving the correct advice. Nonetheless, even when the algorithm is wrong, users agree with the recommendations more than half the time.
\subsubsection{Effect of news topics} %\hspace*{\fill}

% \begin{figure}[t]
%   \centering
%   \includegraphics[width=0.65\textwidth]{user-agreement-per-topic-v2.png}
%   \caption{Comparison of user agreement per topic.}
%   \label{fig:agreement-by-topic}
%   \vspace{-0.5cm}
% \end{figure}

Finally, the percentage agreement in specific news topics can be seen in Figure~\ref{fig:topic-accuracy-agreement} (right). The difference between topics was found to be significant (Friedman test, $p<0.01$) Therefore, it is possible that users are more likely to follow more the algorithm in certain topics than in others. However, similar to our analysis of the topics in user accuracy, these results may require further studies with a greater variety of news articles per topic.

% Also, in this analysis, 30\% of the articles affected by the system incorrectness were not considered because the system incorrectness was not balanced between topics.

% with a p-value much lower than 1\% in the Friedman test between the topics.

\section{Limitations}
Participants recruited through Amazon Mechanical Turk have known limitations, such as perverse incentives to complete the task as quickly as possible without regard to correctness. We attempted to mitigate these by setting a high threshold for ratings when selecting participants, and instituting time checks and analysis of response distributions to ensure that participants were not providing random responses.

Our analysis of topics was post-hoc; we did not intend to compare differences between topics from the outset, thus our dataset was too small to include many instances of each topic. Future work could continue along the lines suggested in our preliminary analysis by expanding the dataset greatly, to create representative sets for different topics.

This study did not cover getting qualitative insights from the users. This feedback could be important to understand users' opinions of the system warnings and the nature of their mental processes when critically evaluating the truthfulness of an article in the context of a system warning (which may be wrong). Future studies could gather more qualitative data and contrast them with quantitative findings.

Some other variables could be modified as well to analyse their impact. We fixed the algorithm accuracy of 70\%; future studies could test the impact of different levels of accuracy. We did not vary the formatting of the articles or warnings, but future studies might explore showing warnings with different text messages. Possible changes to the text messages include stating different verbs for the warning "Our system believes...", for example, "classifies" or "thinks", or changing the wording overall. Article attributes could also be changed to give the users more information, such as showing the news source in the extract. Future studies may explore the interaction of these presentational factors with algorithm warnings.

\section{Discussion}
We find that a fake news detection algorithm, even one that is only 70\% accurate, can help users distinguish fake and real news. Overall, showing warnings helped improve user accuracy by 7.5 percentage points, and this improvement was statistically significant.

User accuracy in detecting fake news was greater than detecting real news. This may be deception bias, which has also been found in other studies \cite{luo_credibility_2020}, where users tend to be more skeptical when their attention is drawn to the possibility of encountering false articles. Fake news article headlines tend to be less subtle, which might make them easier to identify.

In contrast to Snijders et al., \cite{snijders2022humans}, our study measured a larger improvement in accuracy when using algorithmic advice. This is likely due to multiple methodological differences. Our design has greater ecological validity: our participants did not choose when to receive advice, they were not shown the ground truth in order to help calibrate trust in the model, and they did not see the same article repeatedly and therefore could not retroactively update their judgement after having seen the ground truth.

% We believe our study benefited from a more practical method where the user does not choose when to receive advice; he is only exposed to a specific article once and is not exposed to results during the survey, reducing the risks of other factors influencing their decision.

At the time of this study, we inhabit an information culture where most web users are not consciously critical of the truth of articles in their day-to-day reading. To increase the ecological validity of our findings, a future study could observe people's behaviour in a more naturalistic, day-to-day setting. For example, this might take the form of a longitudinal diary study, complemented with logs of their interactions in a social media platform or news website.

% , where users may not focus on the articles in detail and analyse if there exist differences in accuracy and trust.

Algorithm correctness was also a significant factor. Users agreed with the algorithm's advice 72.5\% of the time. Even when the algorithm gave wrong advice, users agreed with their recommendations more than half the time. When the system was correct, users had higher accuracy than when the system was wrong; possibly indicating an unwillingness to go against the algorithm's decision, which can be viewed as ``overtrust'' or ``inappropriate trust'' \cite{yang2020visual}. For example, four users followed the algorithm's decision 100\% of the time, placing total trust in the algorithm. These results suggest that it would be better not to give a warning, than to give an incorrect warning. At a high level of trust, increasing the algorithm's accuracy is also likely to increase users' accuracy.

% Additionally, as a measure of users' trust in the algorithm, u

% The topics of the news were also found to be significant in the results of user accuracy and user agreement. This could be due possibly to the user's previous knowledge and beliefs. Due to the limited amount of news, it was not possible to do a full-scale analysis of how different controversial topics might impact users' trust and accuracy. Further experiments could explore this if a more extensive dataset is used.

\section{Conclusion}
Even though the performance of fake news detection algorithms has improved in recent research, and they are essential tools in the fight against misinformation, they are still not (and may never be) fully accurate. It is therefore necessary to evaluate the effects of using an imperfect algorithm on users' trust, to test whether such systems are worth using despite their possibility for error.

% Users may not always follow the algorithm's advice, and algorithmic errors can have consequences on users' choices when browsing online. 
In our study, the user accuracy in classifying news as real or fake increased by 7.5 percentage points when assisted by an algorithm that itself was only 70\% accurate. This accuracy can be affected by other factors such as the nature of the article, the algorithm correctness and the news topics. 

We further found that user agreement with the algorithm was 78.9\% when the algorithm was correct, and 57.8\% when it was wrong. A high level of user trust in the algorithm, even when incorrect, indicates that no warnings may be better than incorrect warnings, and that any improvement in the accuracy of the system should translate directly into improvement in the user's ability to detect fake news.

%Bibliography
\bibliographystyle{unsrt}  
\bibliography{references}

\end{document}